\documentclass[aps,prb,reprint,showpacs,superscriptaddress,longbibliography]{revtex4-2}
\usepackage{mathtools,amssymb,graphicx,units}

\usepackage[plainpages=false,pdfpagelabels,colorlinks=true,linkcolor=red,urlcolor=blue,citecolor=blue,pdftitle={Title},pdfauthor={},pdfdisplaydoctitle=true,pdfduplex=DuplexFlipLongEdge]{hyperref}
\usepackage{verbatim}
\usepackage[english]{babel}
\usepackage{color}

\usepackage{ulem}

\begin{document}
\title{Higher-order topological insulator in a modified Haldane-Hubbard model}
\author{Tian-Cheng Yi}

\affiliation{Beijing Computational Science Research Center, Beijing 100193, China}

\author{Hai-Qing Lin}
\affiliation{Beijing Computational Science Research Center, Beijing 100193, China}
\affiliation{School of Physics, Zhejiang University, Hangzhou, 310058, China}
\author{Rubem Mondaini}
\email{rmondaini@csrc.ac.cn}
\affiliation{Beijing Computational Science Research Center, Beijing 100193, China}

\begin{abstract}

We investigate the ground-state phase diagram of a modified spinless Haldane-Hubbard model, with broken threefold rotational symmetry, employing exact diagonalization calculations. The interplay of asymmetry, interactions, and topology gives rise to a rich phase diagram. The non-interacting limit of the Hamiltonian  exhibits a higher-order topological insulator characterized by the existence of corner modes, in contrast to known chiral edge metallic states of the standard Haldane model. Our investigation demonstrates that these symmetry-protected states are robust to the presence of finite interactions. Furthermore, in certain regimes of parameters, we show that a topological Mott insulator exists in this model, where a non-trivial topological bulk coexists with an interaction-driven charge-density-wave, whose emergence is characterized by a $Z_2$-symmetry breaking within the 3$d$-Ising universality class.
\end{abstract}

\maketitle

\section{Introduction}

The discovery of topological insulators (TIs) and superconductors have attracted considerable attention, being extensively investigated in many different systems in recent years, such as in electronic~\cite{RevModPhys.82.3045, RevModPhys.83.1057} or photonic systems~\cite{RevModPhys.91.015006}. It has culminated with an overall scheme for classifying topological quantum matter depending on the symmetries of the related models~\cite{RevModPhys.88.035005}. A distinctive feature of TIs is the bulk-boundary correspondence: a $d$-dimensional TI exhibits topologically protected gapless states on its $(d-1)$-dimensional boundaries, mapped by the existence of a finite topological invariant in the bulk. 

Even more recently, it has been noticed that in some cases, the protected modes are instead restricted to hinges or corners of the system~\cite{benalcazar2017quantized, schindler2018higher}. This gave rise to the concept of higher-order topological insulators (HOTIs), which exhibit gapped $(d-1)$-dimensional boundaries while supporting gapless, topologically protected states on a lower $(d-n)$-dimensional boundary for $n \geqslant 2$. HOTIs have been experimentally observed in materials~\cite{Schindler2018expt,Yue2019} or emulated in engineered platforms as mechanical or photonic metamaterials~\cite{Serra-Garcia2018, Li2020}, or electric~\cite{Imhof2018, Zhang2021, Lv2021} and resonator circuits~\cite{Peterson2018, Xue2019, Ni2019, Yamada2022}. Theoretically, they have been classified in a variety of Hamiltonians, including modified versions of celebrated models such as the Su-Schrieffer-Heeger~\cite{luo2022higher}, Aubry-Andr{\'e}-Harper~\cite{zeng2020higher}, and Haldane models~\cite{Wang2021}. 

The latter~\cite{Haldane1988} realizes a quantum anomalous Hall insulator featuring a topologically protected chiral state carrying a dissipationless current without an external magnetic field~\cite{Liu2016, Chang2022}. It has been generalized to understand the effects of interactions~\cite{Varney2010, Wang2010, Varney2011, Vanhala2016, Imriska2016, Tupitsyn2019, Shao2021, Mai2022}, disorder~\cite{Sriluckshmy2018, Goncalves2018}, and their combined interplay~\cite{Yi2021} in its topological properties. A common theme in these results is that, at half-filling, once interactions are sufficiently large to induce a finite local order parameter, protection of edge modes is absent, and trivial insulating phases ensue.

In the context of higher-order topology, decorated tight-binding models in the honeycomb lattice have been shown to support protected corner modes~\cite{Xue2021}. In particular, a generalization of the Haldane model that breaks its three-fold rotational symmetry ${\cal C}_3$ leads to a HOTI, provided that inversion symmetry is preserved~\cite{Wang2021}. Physically, this modification corresponds to a type of uniaxial strain in the system~\cite{Pereira2009}, whose characterization of first-order topological properties have been performed earlier~\cite{Ho2017, Mannai2020}.

The combination of these venues, interactions, and high-order topology has been much less explored. For example, recent results in electronic spinful Hamiltonians have demonstrated the formation of gapless, topologically protected spin excitations in corners of the lattice, whereas charge excitations are still gapped~\cite{Kudo2019, Otsuka2021}; formation of gapless corner modes are also seen in either spin models~\cite{Dubinkin2019, Guo2022} or extensions of the Bose-Hubbard model~\cite{Bibo2020}. Yet, observation of interacting fermionic models featuring corner modes related to gapless charge excitations is currently lacking.

\begin{figure*}[t!]
\centering
\includegraphics[width=0.55\columnwidth]{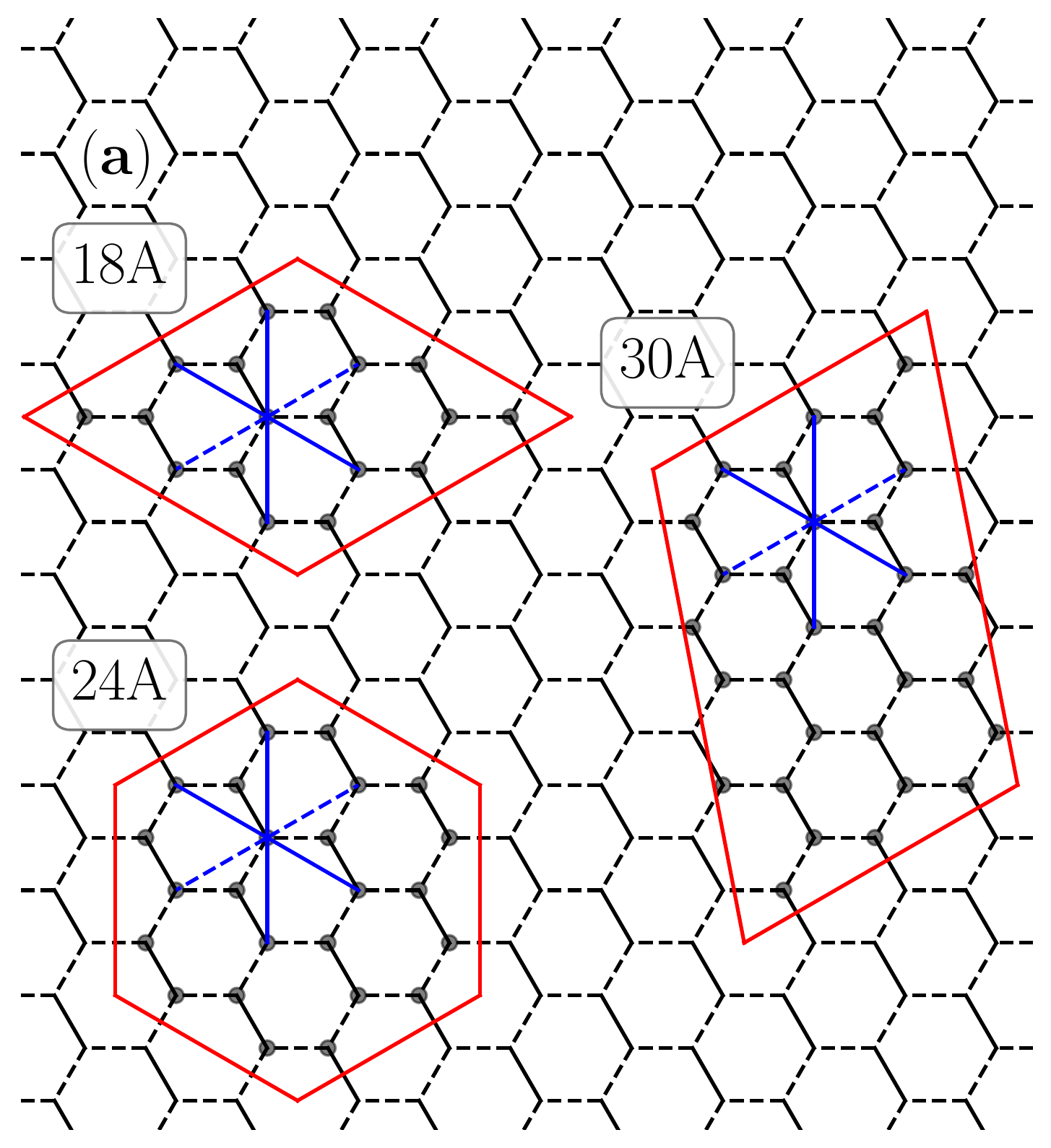}
\includegraphics[width=0.7\columnwidth]{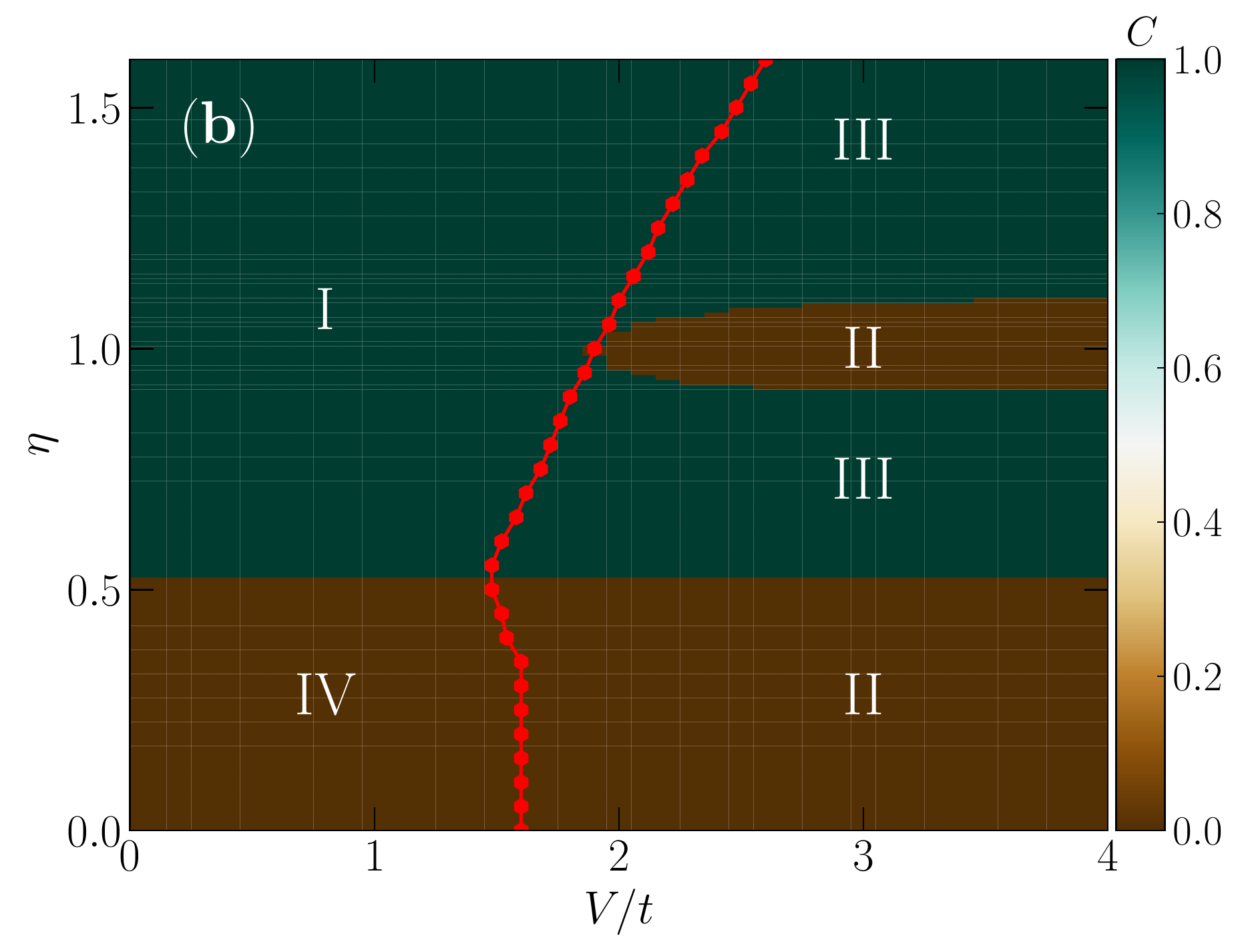}
\includegraphics[width=0.7\columnwidth]{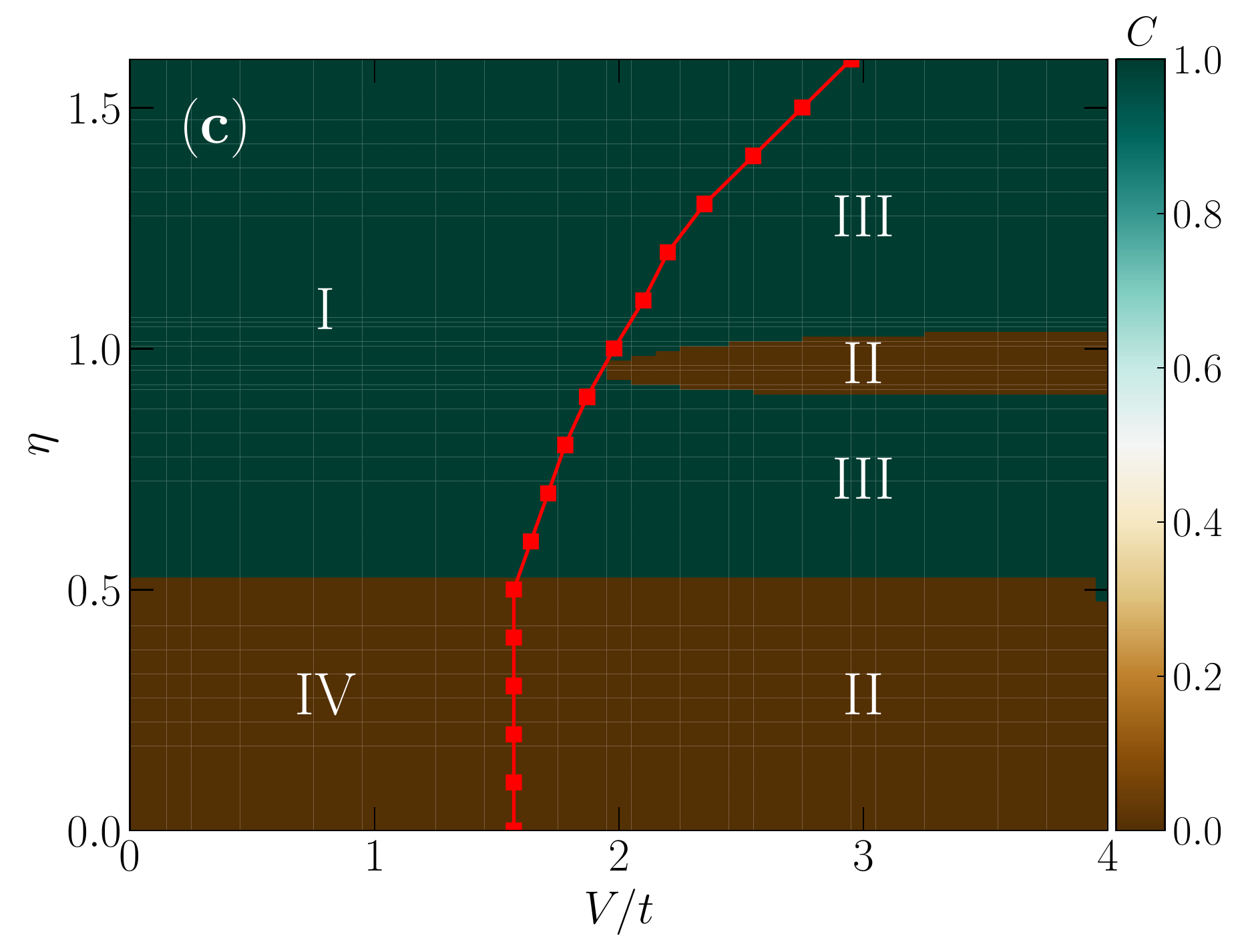}
\caption{(a) Schematic representation of the lattice clusters used in the ED calculations: 18A, 24A, and 30A. Solid and dashed lines represent the bonds for different hopping amplitudes between the nearest (black) and next-nearest (blue) neighbor sites. For clarity, the latter is just depicted for a single site in each cluster. Red solid lines give the cluster boundaries; open and periodic boundary conditions are used in the calculations. (b) and (c) show the phase diagrams obtained for clusters 24A and 30A, respectively. The colors give the value of the (quantized) Chern number, and the connected markers the location of the CDW transition marked by the fidelity susceptibility $\chi_F$. Different phases numbered I to IV, are labeled in the text. In particular, phase IV is the interacting HOTI, adiabatically connected to its $V=0$ counterpart~\cite{Wang2021}.}
\label{fig:Fig1}
\end{figure*}

To fill this gap, we investigated a dimerized Haldane-Hubbard model, whose non-interacting counterpart was introduced in Ref.~\cite{Wang2021}, to understand the robustness of second-order topological properties upon the inclusion of interactions. While many studies have classified the topological properties of the Haldane-Hubbard model~\cite{Varney2010, Wang2010, Varney2011, Vanhala2016, Imriska2016, Tupitsyn2019, Shao2021, Mai2022}, all these cases are limited to the first-order topology, i.e., to the appearance of protected \textit{edge} states in certain regimes of parameters. By using exact diagonalization methods, we uncover the presence of an interacting HOTI, characterizing the formation of protected \textit{corner} modes.

As a by-product of our analysis, we demonstrate the existence of a genuine topological Mott insulator (TMI) in this model~\cite{Rachel2018}, that is, a regime in the parameters in which both insulating behavior stemming from increasing interactions and non-trivial topological properties in the ground-state concomitantly occur. TMIs have a long history of research with often conflicting results: The existence of interaction-driven TMIs has been previously argued~\cite{Raghu08, Wen10, Budich12, Dauphin12, Weeks10, LeiWang12, Ruegg11, Yang11, Yoshida14}, while  other results using unbiased methods~\cite{Garcia_Martinez13, Daghofer14, Motruk15, Capponi15, Scherer15} dispute some of the claims. An exception being two-dimensional systems featuring quadratic band-crossings and weak interactions~\cite{Sun09, Murray14, Venderbos16, Wu16, Zhu16, Vafek10, Wang17}, but here such characterization is unambiguous. On top of that, since the onset of the TMI is manifested by the emergence of charge order, we show that the zero-temperature transition belongs to the 3$d$-Ising universality class, owing to the $Z_2$-symmetry breaking of the corresponding order parameter.

\section{Model and quantities} 
\label{sec:Model}
We consider a modified Haldane-Hubbard model on a honeycomb lattice with Hamiltonian
\begin{eqnarray} 
{\cal \hat H}
&=&-\sum_{{\langle i,j\rangle}} t_1^{ij}\left( \hat c^\dag_i \hat c^{\phantom\dag}_j + {\rm H.c.} \right)
-\sum_{{\langle\langle i,j\rangle\rangle}}
t_2^{ij}
\left(e^{{\rm i} \phi_{ij}} \hat c^\dag_i \hat c^{\phantom \dag}_j + {\rm H.c.} \right)
\nonumber\\&&+\Delta \sum_i (-1)^{i} \hat n_i^{\phantom\dag}
+V \sum_{{\langle ij\rangle}} \hat n_i \hat n_j.
\label{eq:Hamiltonian}
\end{eqnarray}
Here, $\hat c_i^\dag$ ($\hat c_i^{\phantom\dag}$) is the fermion creation (annihilation) operator at site $i$, and $\hat n_i =\hat c^\dag_i \hat c_i^{\phantom\dag}$ is the corresponding number operator. $t_1^{ij}$ ($t_2^{ij}$) gives the nearest-neighbor (next-nearest-neighbor) hopping amplitude, and $\Delta$ is the staggered potential responsible for breaking the symmetry between the two sublattices of a honeycomb lattice. The next-nearest-neighbor hopping term has a complex phase $\phi_{ij} = +\phi (-\phi)$ for  counter-clockwise (clockwise) hoppings and $V$ describes the magnitude of a repulsive nearest-neighbor interaction. 

When $t_1^{ij}=t_1$, $t_2^{ij}=t_2$, ${\cal \hat H}$ simplifies to the original homogeneous spinless Haldane-Hubbard model~\cite{Varney2010, Varney2011}. Alternatively, one can introduce a dimerization of both nearest and next-nearest neighbor hopping amplitudes along one preferential direction, as schematically represented in Fig.~\ref{fig:Fig1}(a). These establish two types of bonds for each hopping term, whose amplitude is given respectively by $t_{1s, 2s}$ and $t_{1d, 2d}$. 

By defining the dimerization strength, $\eta\equiv t_{1d}/t_{1s}\equiv t_{2d}/t_{2s}$, Ref.~\cite{Wang2021} showed that if departing from the homogeneous case with $\eta = 1$, a gap closes and reopens at $\eta = \pm 0.5$, such that for values $0.5>\eta>-0.5$ the non-interacting ($V=0$) Hamiltonian characterizes a HOTI, provided inversion symmetry is preserved (i.e., with $\Delta = 0$). This phase is described by having an associated zero value of the Chern number $C$, while yet  harboring corner modes. Such a topological invariant is computed via the integration of the Berry curvature~\cite{Niu1985},
\begin{equation}
    C = \int \frac{d\phi_x d\phi_y}{2 \pi { \rm i}} \left( \langle\partial_{\phi_x}
      \Psi_0^\ast | \partial_{\phi_y} \Psi_0\rangle - \langle{\partial_{\phi_y}
      \Psi_0^\ast | \partial_{\phi_x} \Psi_0\rangle} \right),
      \label{eq:Chern}
\end{equation}
after introducing twisted boundary conditions $\{\phi_x,\phi_y\}$~\cite{Poilblanc1991} when obtaining the ground-state $|\Psi_0\rangle$ of ${\cal \hat H}$. In practice, a sufficiently discretized version of Eq.~\eqref{eq:Chern} suffices~\cite{Fukui05}, as has been shown  to converge to a quantized Chern number in the same model~\cite{Varney2011, Shao2021, Yi2021}.
 

In what follows, we investigate the low-lying spectral properties of ${\cal \hat H}$ by using exact diagonalization (ED) in finite clusters ranging from $N_s = 18$ to 30 sites, focusing the investigation at half-filling, i.e., $N_e\equiv \sum_i \langle \hat n_i\rangle = N_s/2$. A representation of the clusters used, all featuring the $K$ high-symmetry point as a valid momentum value, is given in Fig.~\ref{fig:Fig1}. Solid and dashed lines describe the different hopping amplitudes. When $\eta < 1$, they represent, respectively, the strong and weak bonds. For clarity, the (blue) solid and dashed lines denoting dimerized hopping amplitudes between next-nearest neighbor sites are expressed on a single site for each cluster; red lines give the cluster boundaries, which are chosen to be either open or periodic, depending on the quantities one is interested in.

On top of the topological properties, we characterize the formation of a charge-density wave (CDW) associated with a trivial Mott insulating (MI) behavior at sufficiently large $V$. For that, we quantify the ${\bf k}=0$ CDW structure factor~\cite{Varney2010, Varney2011, Shao2021, Shao2021b, Yi2021}%
\begin{align}
  \label{eqn:struct}
  S_{\rm CDW}  &\equiv \frac{1}{N}\sum_{i,j} C({\bf r}_i - {\bf r}_j)
\end{align}
with density correlations
\begin{align}
  C({\bf r}_i - {\bf r}_j) &= \langle (\hat n_i^a - \hat n_i^b) (\hat n_j^a - \hat n_j^b) \rangle,
\end{align}
where $\hat n_i^a$ and $\hat n_i^b$ are the number operators on sublattices $a$ and $b$
in the $i$-th unit cell, respectively, and $N = N_s/2$ is the total number of unit cells. This quantity is extensive in $N$ once a long-range CDW order sets in, tracking thus the formation of the corresponding local order parameter.

Lastly, we further quantify the fidelity susceptibility~\cite{Zanardi06, CamposVenuti07, Zanardi07, You2007},
\begin{equation}
\chi_{F} = \frac{2}{N_s} \frac{1 - \langle \Psi_0(V)|\Psi_0(V+dV)\rangle}{dV^2},
\end{equation}
with $dV = 10^{-3}$ in our calculations, which identifies a continuous quantum phase transition through the location of an extensive peak in $N_s$ in the region parameters of interest~\cite{Yang07, Varney2010, Jia11, Mondaini15, Jin2022}. In the case of first-order topological phase transitions, as the CI-MI for $\eta = 1$, the fidelity susceptibility exhibits discontinuities so long as the lattice possesses the corresponding high-symmetry point where the closing of the excitation gap occurs~\cite{Varney2010, Varney2011, Shao2021}. 

Immediate verification of a first-order phase transition, invariably tied to the modification of the topological invariant, is obtained by the computation of the excitation (or many-body) gap:
\begin{eqnarray}
    \Delta_{\rm{m}} = E_1(N_s/2)- E_0(N_s/2),
\end{eqnarray}
which quantifies the energy difference between the two-lowest eigenvalues of Eq.~\eqref{eq:Hamiltonian} for the studied filling factor $N_e = N_s/2$.

Other quantities, such as charge compressibilities and charge gaps, mainly used to identify the protected corner modes, are defined subsequently in the corresponding sections. In what follows, we choose $t_{2s}/t_{1s} = 0.2$, $\phi = \pi/2$ and establish $t_{1s}=t$ as unit of energy. We also focus on the case that preserves inversion symmetry, i.e., $\Delta=0$, since that is a precondition for the manifestation of a HOTI in this model~\cite{Wang2021}. Lastly, we narrow the investigation to the $\eta>0$ regime.

\section{Results}

\subsection{Phase diagram}

We start by describing the ground-state phase diagram of Eq.~\eqref{eq:Hamiltonian} in the space of parameters $\eta-V$ in Fig.~\ref{fig:Fig1}(b) and \ref{fig:Fig1}(c), for the two largest clusters we study, 24A and 30A, respectively, using PBCs. We note the existence of four different phases based on the analysis of the quantized Chern number and the peak location of the fidelity susceptibility, which tracks the onset of CDW order when increasing the interaction strength:
\begin{itemize}
\item[(I)] topological (Chern) insulator ($C=1$, no CDW order);
\item[(II)] Mott insulator ($C=0$, CDW order);
\item[(III)] topological Mott insulator ($C=1$, CDW order);
\item[(IV)] higher-order topological insulator ($C=0$, no CDW order).
\end{itemize}
In the absence of dimerization, $\eta=1$, the transition between phases I and ${\rm II}$ describes the known results of the ${\cal C}_3$-symmetric Haldane-Hubbard model, namely, that a finite local order parameter is incompatible with non-trivial topology at half-filling~\cite{Varney2010,Varney2011}. That is, at a critical $V=V_c\simeq 2t$, the Chern insulator (I) gives way to a trivial Mott insulator (II). We further notice that this phase II defines a lobe that narrows in its $\eta$-support when increasing the system size. While it is unclear whether phase II will be constrained to the $V>V_c$, $\eta=1$ line in the thermodynamic limit, this trend makes apparent the robustness of phase III, namely, a non-trivial insulator ($C=1$) that also exhibits CDW order. As far as we know, this is the first evidence of a topological Mott insulator in the Haldane-Hubbard model using unbiased methods.

For phase IV, we note that the non-interacting HOTI with $|\eta|<0.5$~\cite{Wang2021} adiabatically connects to its interacting counterpart. As expected, it exhibits a $C=0$ topological invariant, but past $V/t \gtrsim 1.6$ is replaced by another trivial (i.e. $C=0$) phase, which instead displays CDW order. This is again a manifestation of phase II, a trivial Mott insulator. Still, unlike in the case of $\eta=1$, since there is no change of the topological invariant, the transition is second-order, with a spontaneous $Z_2$ symmetry-breaking.

Having presented the main features of the phase diagram, we now discuss details of how the different regions were inferred in the next two subsections.

\begin{figure}[ht!]
\centering
\includegraphics[width=1\columnwidth]{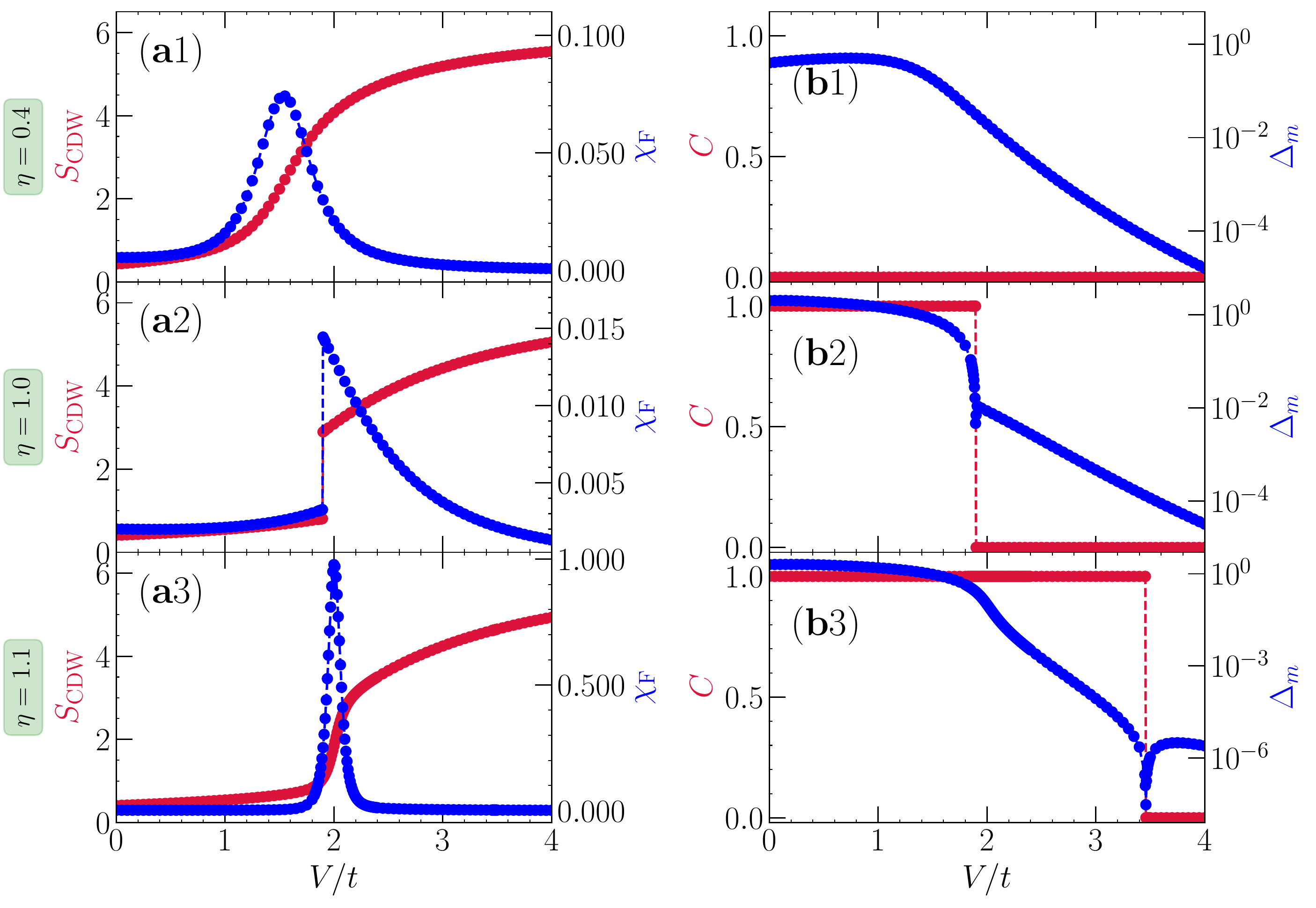}
\caption{(a1)-(a3) CDW structure factor (left) and fidelity susceptibility (right) along three cuts of the phase diagram, with $\eta = 0.4; 1$ and $1.1$. (b1-b3) Corresponding results for the Chern number (left) and the excitation gap (right). Here we use the 24A cluster with PBCs.
}
\label{fig:Fig2_24A_cuts}
\end{figure}

\subsection{Topological and Mott transitions}
A remarkable feature that our phase diagram [Fig.~\ref{fig:Fig1}] exposes is that the topological phase transition is not necessarily accompanied by a Mott one. Only in the case of the homogeneous (i.e., $\eta=1$) Haldane-Hubbard model does this hold~\cite {Varney2010,Varney2011}. Finite hopping-dimerization breaks such constraint, and Fig.~\ref{fig:Fig2_24A_cuts} displays such dissociation. If we define the critical interaction that triggers a topological (Mott) transition at a given $\eta$ as $V_T$ ($V_M$), a known first-order phase transition at $V_T=V_M$ occurs in the homogeneous case, marked by a simultaneous discontinuity of the structure factor and the fidelity susceptibility [Fig.~\ref{fig:Fig2_24A_cuts}(a2)]; also accompanied by a change of the Chern number, expressed by the closing of the excitation gap [Fig.~\ref{fig:Fig2_24A_cuts}(b2)]. In turn, if $\eta\neq 1$, $V_T\neq V_M$ in general. The topological transition is still marked by the location where $\Delta_{\rm m} = 0$ [e.g., see Fig.~\ref{fig:Fig2_24A_cuts}(b3)], which is no longer related to the point at which $\chi_F$ displays a peak [Fig.~\ref{fig:Fig2_24A_cuts}(a3)]. Finite-size effects are addressed in Appendix~\ref{sec:FSE}, showing that these results are qualitatively unchanged for other lattice sizes.

Since the Mott transition independently occurs from the topological character of the ground state change, it can now reflect its typical second-order nature. The fidelity susceptibility becomes continuous, as is the CDW structure factor, where the ensuing charge ordering breaks a $Z_2$ symmetry. Consequently, this zero-temperature phase transition belongs to the (2+1)-$d$ Ising universality class and $S_{\rm CDW}$ should obey the following scaling ansatz:
\begin{equation}
    N_s^{-\gamma/2\nu}S_{\rm CDW} = g[(V-V_M^c)N_s^{1/2\nu}]\ .
    \label{eq:scaling}
\end{equation}
In such universality class, the exponent $\nu$ related to the divergence of the correlation length is $\nu = 0.629971(4)$ while $\gamma$, related to the singular behavior of two-point correlation functions, is $\gamma = 1.237075(10)$~\cite{Pelissetto2002}. 

\begin{figure}[ht]
\centering
\includegraphics[width=0.99\columnwidth]{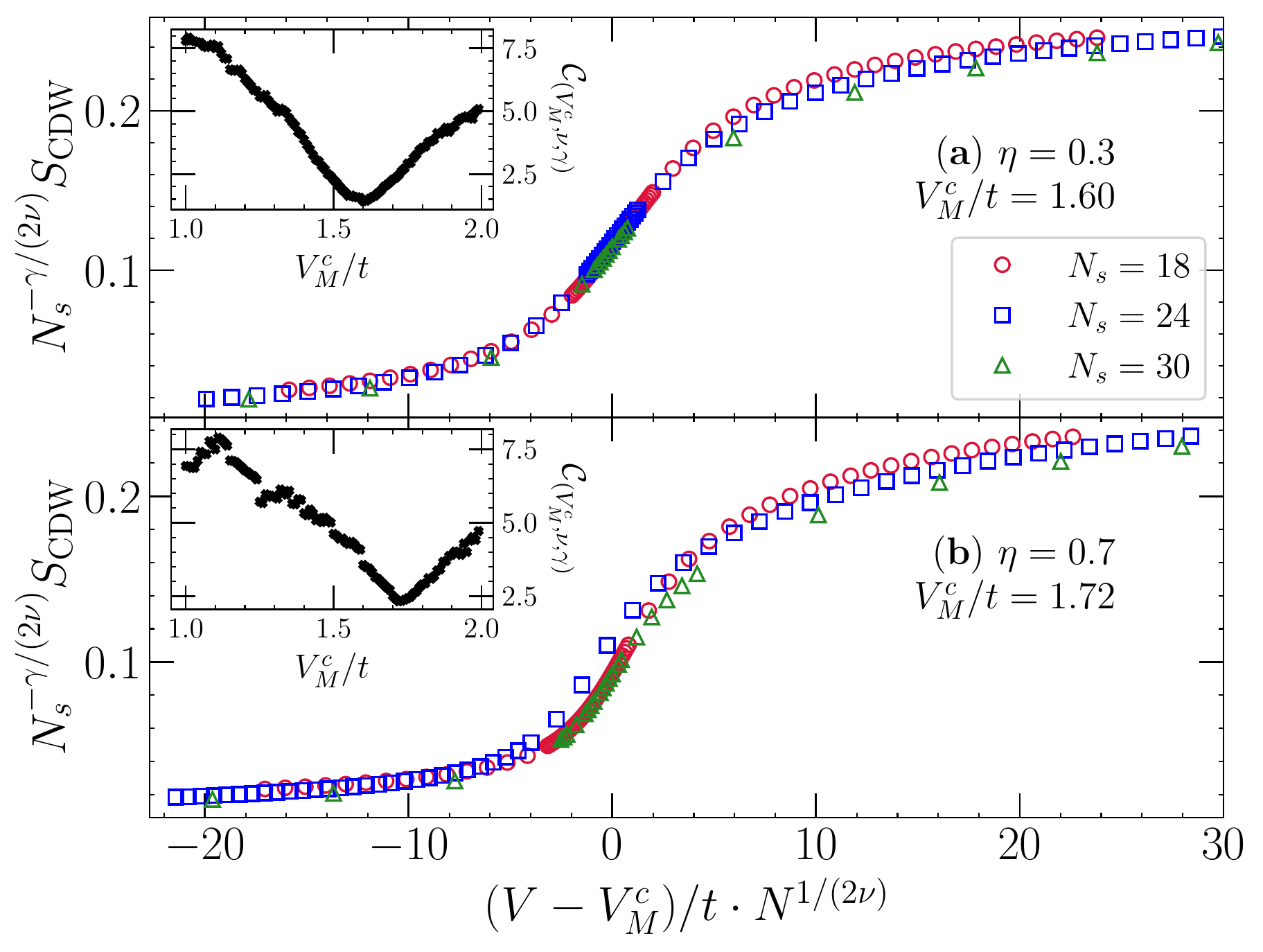}
\caption{Scaling behavior of $S_{\rm{CDW}}$ according to the scaling ansatz [Eq.~\eqref{eq:scaling}] with 3$d$-Ising exponents $\gamma = 1.237075$ and $\nu = 0.629971$. (a) $\eta = 0.3$, describing the Mott transition within the $C=0$ regime; (b) $\eta = 0.7$, Mott transition within the $C=1$ phase. The insets show the corresponding values of the cost function (see text) for the scaling collapse as a function of $V_M^c$.}
\label{fig:cdw_scaling}
\end{figure}

Figure \ref{fig:cdw_scaling} shows the scaling analysis using the three cluster sizes available and values of $\eta = 0.3$ and $0.7$. We estimate the critical interaction $V_M^c$ in the thermodynamic limit by the minimum value of a cost function that quantifies the scaling collapse. It is written as ${\cal C}_{(V_c, \nu, \gamma)} = (\sum_j |y_{j+1} - y_j|)/(\max\{y_j\} - \min\{y_j\})-1$~\cite{Suntajs2020, Mondaini2022b}, where $y_j$ are the values of $N_s^{-\gamma/2\nu}S_{\rm CDW}$ ordered according to their corresponding $(V-V_M^c)N_s^{1/2\nu}$'s [see insets in Fig.~\ref{fig:cdw_scaling}]. A relatively good collapse and agreement with the expected critical exponents are obtained, despite having a maximal value of the linear lattice size $L = N_s^{1/2} \simeq 5.48$. For $\eta = 0.3$, the Chern number is zero across the whole range of interactions investigated [see Fig.~\ref{fig:Fig1}], reflecting the aforementioned interacting HOTI to trivial Mott insulating transition. With $\eta=0.7$, however, $C=1$ irrespective of the $V$ magnitude studied when entering the topological Mott insulating phase. This indicates that the 3$d$-Ising universality class describes the Mott transition, whether or not the ground state exhibits a trivial Chern number.

\subsection{Many-body gap, charge gap, and the HOTI}

\begin{figure}[ht!]
\centering
\includegraphics[width=0.9\columnwidth]{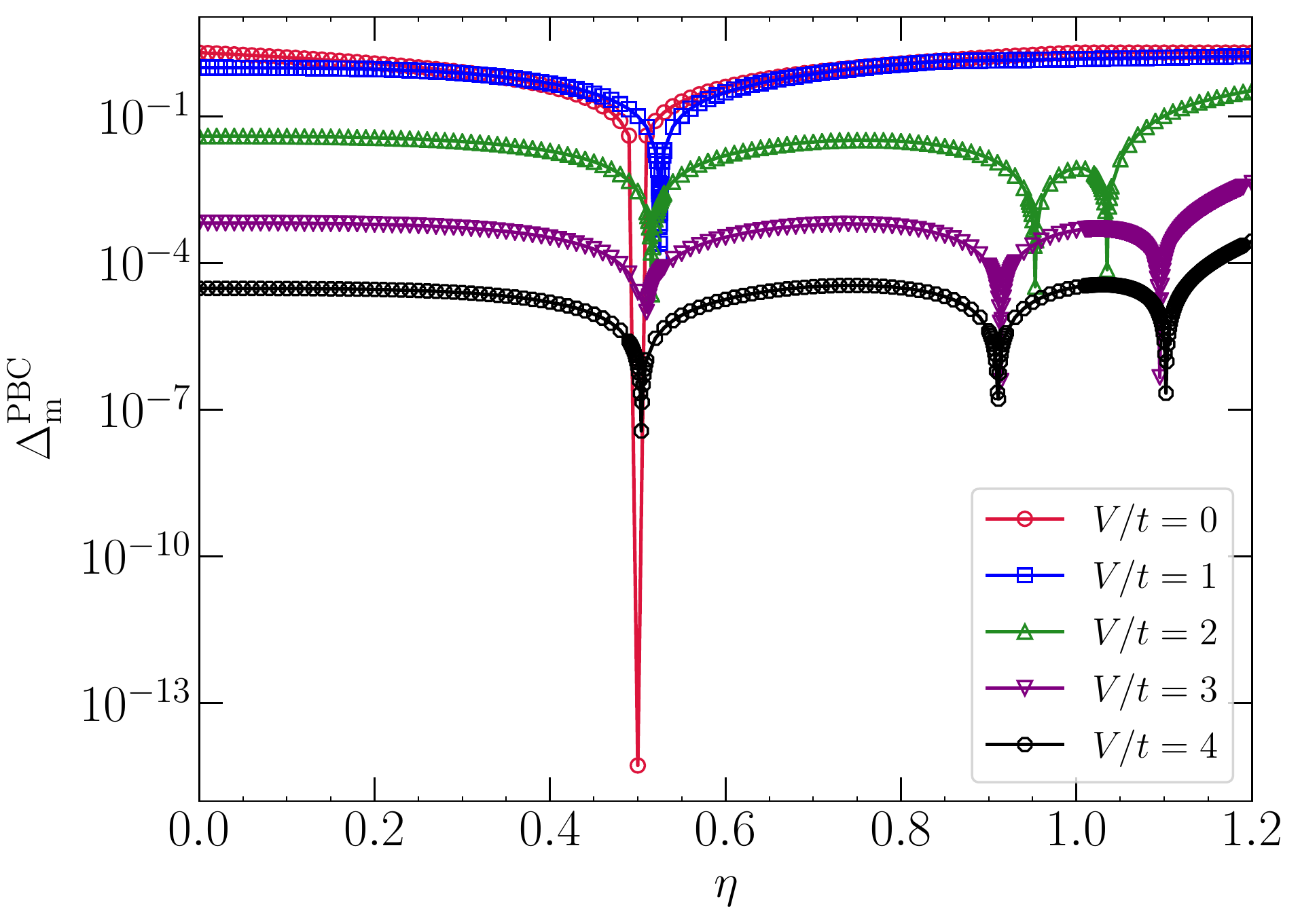}
\caption{Many-body gap $\Delta_{\rm{m}}$ dependence with different dimerization values along several cuts in the phase diagram with fixed interaction strengths. Here the 24A cluster is used with PBCs. The dips point to the critical value of $\eta$, where a bulk topological transition occurs. }
\label{fig:charge_gap_eta}
\end{figure}

As mentioned above, direct evidence of the topological transition involving the change of the corresponding topological invariant is seen via the closing of the excitation gap. Figure~\ref{fig:charge_gap_eta} summarizes the dependence with the dimerization parameter $\eta$ with different interaction magnitudes $V$. The non-interacting limit is well-marked by $\Delta_{\rm m}^{\rm PBC}\to 0$ at $\eta=0.5$, describing the HOTI to CI transition introduced in Ref.~\cite{Wang2021}. Within the interacting regime, various values of $V$ lead to a gap closing occurring roughly at the same location, while if the interactions are sufficiently large, a double dip structure centered around the ${\cal C}_3$-symmetric Haldane-Hubbard model ($\eta = 1$) marks the trivial Mott insulating lobes described in the phase diagram, Fig.~\ref{fig:Fig1}.

Although the excitation gap is sufficient to identify the regimes where the Chern number changes, it completely misses the characterization of the higher-order topological regime. This is evident if spanning the interactions with $\eta <0.5$: $C$ is always zero and $\Delta_{\rm m}$ remains finite across a wide range of interactions [see e.g. Fig.~\ref{fig:Fig2_24A_cuts}(b1)]. Quantification of the HOTI and its corner modes relies thus on other metrics. In particular, in the context of interacting systems, one can no longer refer to such modes as gapless excitations in the single-particle spectrum. Instead, one expects its manifestation via the existence of gapless \textit{charge excitations}. The relevant metric is the charge gap defined as,
\begin{equation}
    \Delta_{\rm{c}} = E_0(N_s/2 + 1) + E_0(N_s/2 - 1) - 2 E_0(N_s/2)\ ,
    \label{eq:m_gap}
\end{equation}
which computes the difference in chemical potentials of adding and removing a single particle upon the half-filling $N_e = N_s/2$ we study. To understand the regimes where the HOTI occurs, we need thus to contrast the charge gaps on clusters employing both PBCs and OBCs. Only in the latter can a possible manifestation of corner modes take place.

\begin{figure}[t!]
\centering
\includegraphics[width=1\columnwidth]{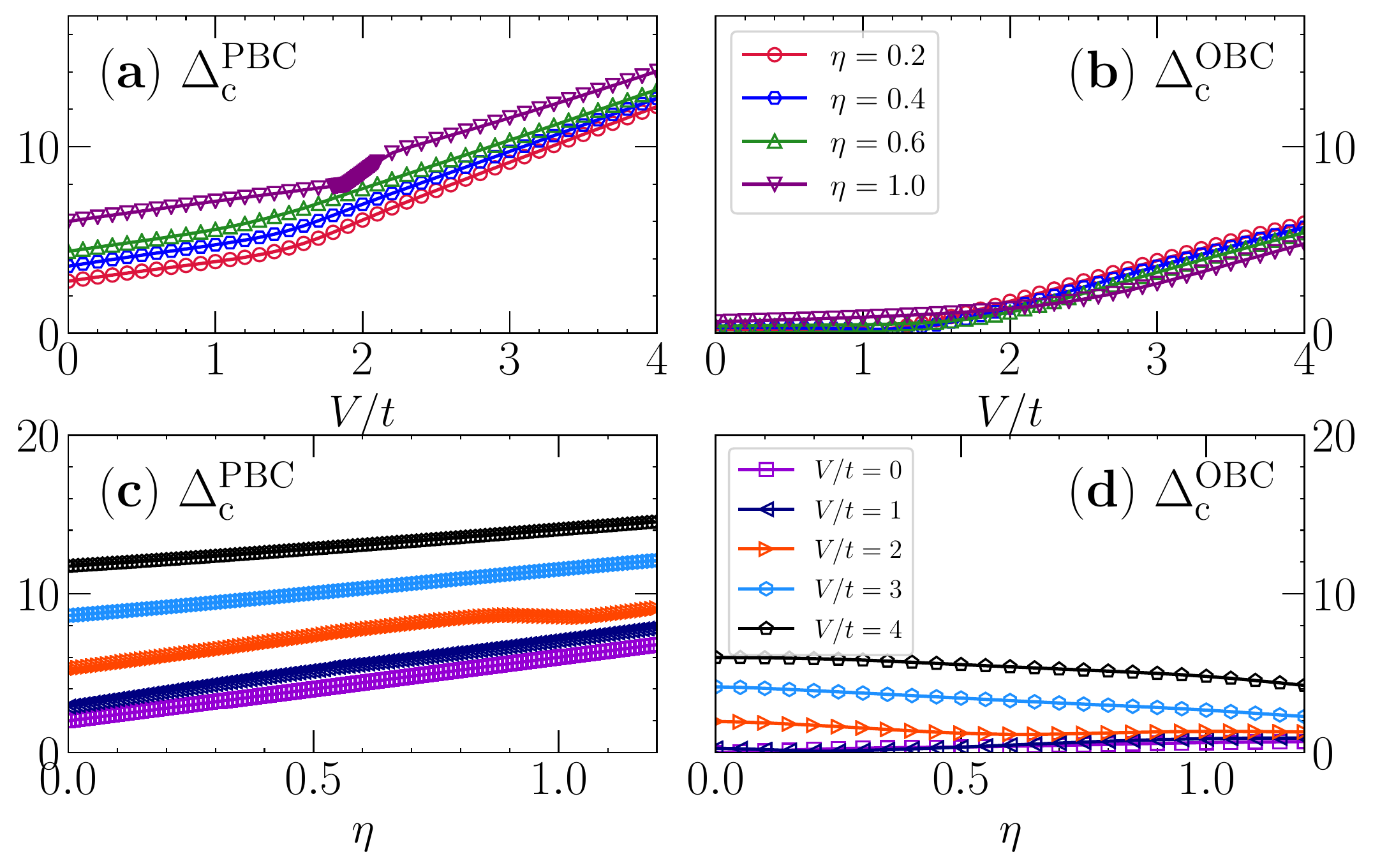}
\caption{Charge gap $\Delta_{\rm c}$ dependence on $V$ [(a) and (b)] and $\eta$ [(c) and (d)], contrasting both PBC (left panels) and OBC (right) in the 24A cluster. While always finite, $\Delta_{\rm c}^{\rm OBC}$ is the smallest when $V\lesssim V_M$ and $\eta \lesssim 0.5$, a regime where the interacting HOTI phase is suggested.}
\label{fig:charge_gap_PBC_OBC}
\end{figure}

Figure~\ref{fig:charge_gap_PBC_OBC} characterizes this on the 24A cluster, showing that (i) for the same parameter's settings, the OBC charge gaps are always smaller than its PBC counterpart, despite being finite due to size effects; (ii) concerning its $V$-dependence, $\Delta_{\rm c}^{\rm OBC}$ only steadily increases when $V\gtrsim V_M$, i.e., when charge ordering sets in, while at $\eta \lesssim 0.5$ and $V\lesssim V_M$ the smallest gaps are obtained -- this is the regime where according to the phase diagram a HOTI manifests.

Although suggestive, such an analysis is superficial in establishing the existence of corner modes. Direct evidence can be put forward by defining the site-resolved compressibility,
\begin{equation}
K_c(i)=
\frac{\partial \langle \hat n_i \rangle}{\partial \mu}
\approx \frac{
\langle \hat n_i \rangle_{N_e+1}
-\langle \hat n_i \rangle_{N_e}
}{\mu_+-\mu_-}\ , 
\label{eq:Kc} 
\end{equation}
where $\mu_+=E_0(N_e+1)-E_0(N_e)$ and $\mu_-=E_0(N_e+1)-E_0(N_e-1)$ are the chemical potentials of adding and removing a single-charge, respectively. Similar analysis has been employed in the context of spin-corner modes in other interacting models exhibiting higher-order topology~\cite{Otsuka2021}. 

\begin{figure}[hb!]
\centering
\includegraphics[width=1\columnwidth]{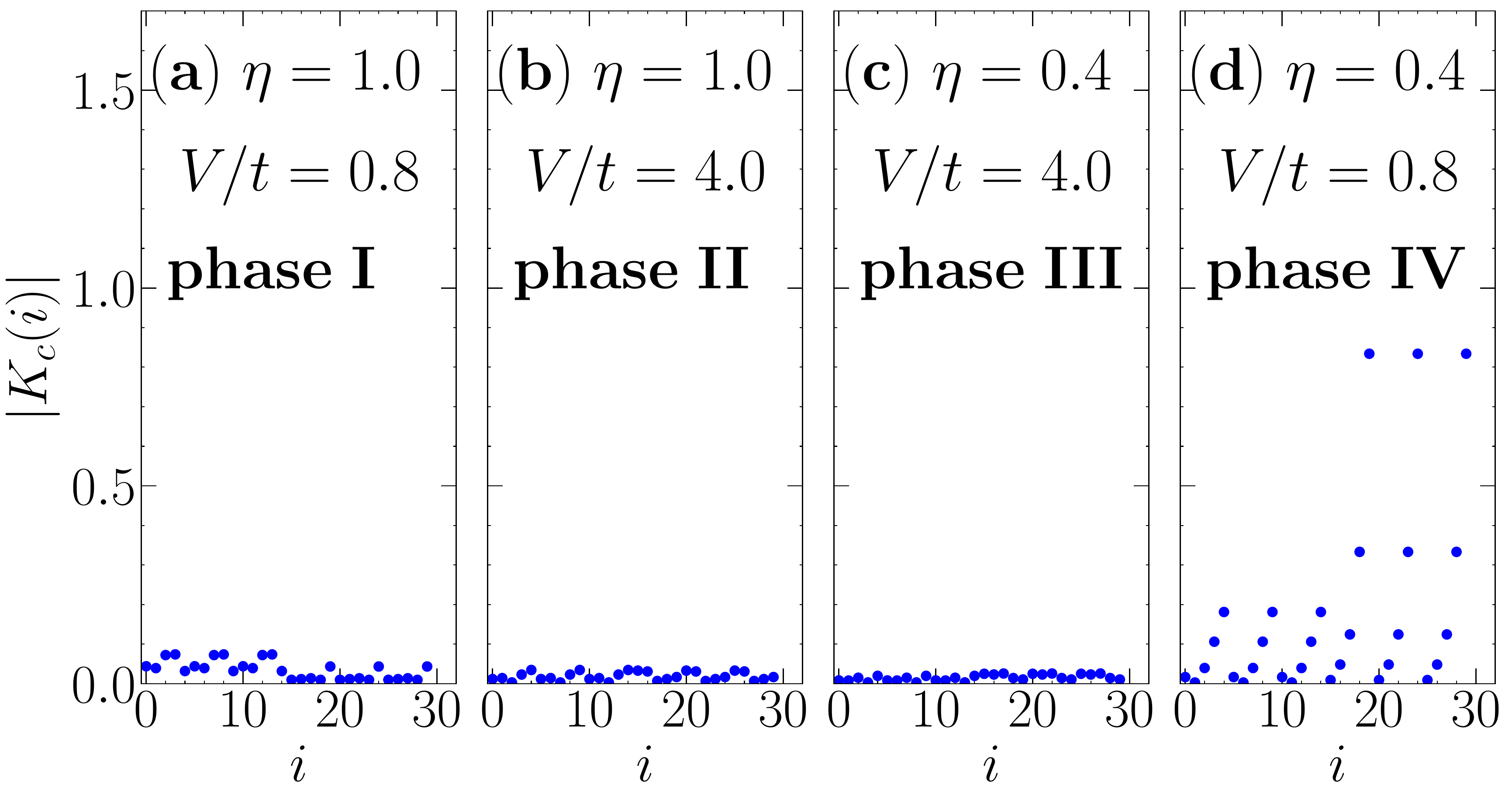}
\caption{The absolute value of the site-resolved charge compressibility in different phases, according to the phase diagram, for the 30A cluster. Here mixed (or cylindrical) boundary-conditions are used (see text) and specific details with increasing interactions $V$ and dimerization $\eta$ can be seen at Figs.~\ref{fig:Kc_eta} and~\ref{fig:Kc_V}.}
\label{fig:Kc}
\end{figure}

We report in Fig.~\ref{fig:Kc} the lattice profile of compressibilities $K_c(i)$ in four representative points of the phase diagram for cluster 30A. While for parameters in phases I, II, and III $|K_c(i)| \simeq 0$ across the whole lattice, in phase IV, it is clear that much higher compressibilities are obtained for selected sites. This is direct evidence of the interacting HOTI in the modified Haldane-Hubbard model.

\begin{figure}[t!]
\centering
\includegraphics[width=0.8\columnwidth]{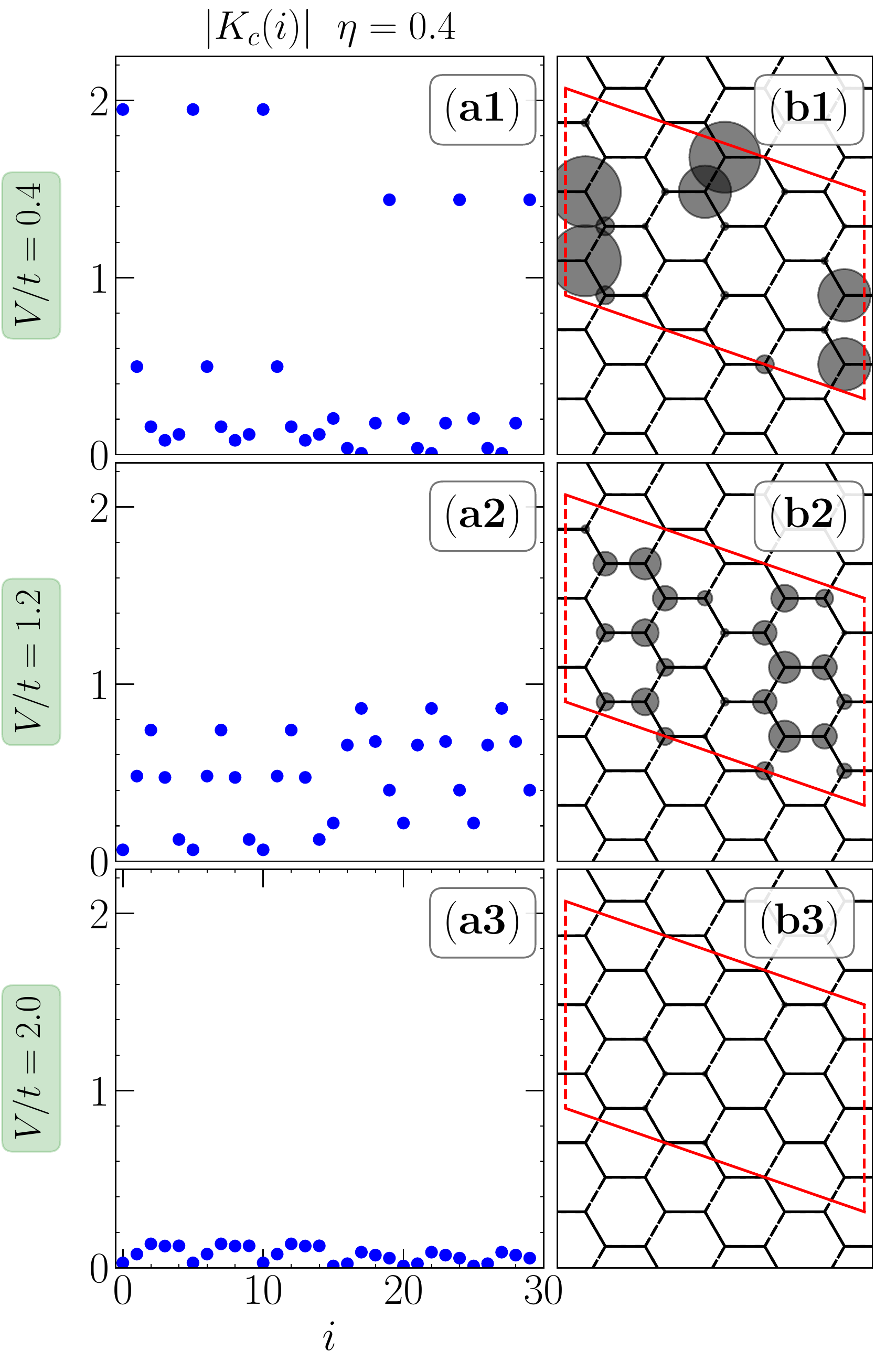}
\caption{The absolute value of the site-resolved charge compressibility with increasing interactions, $V/t = 0.4, 1.2$ and 2.0, in panels (a1)-(a3), with fixed $\eta = 0.4$. Panels (b1)-(b3) show the same data but are represented as markers in the 30A cluster whose size is proportional to the $|K_c(i)|$ value.}
\label{fig:Kc_eta}
\end{figure}

\begin{figure}[t!]
\centering
\includegraphics[width=0.8\columnwidth]{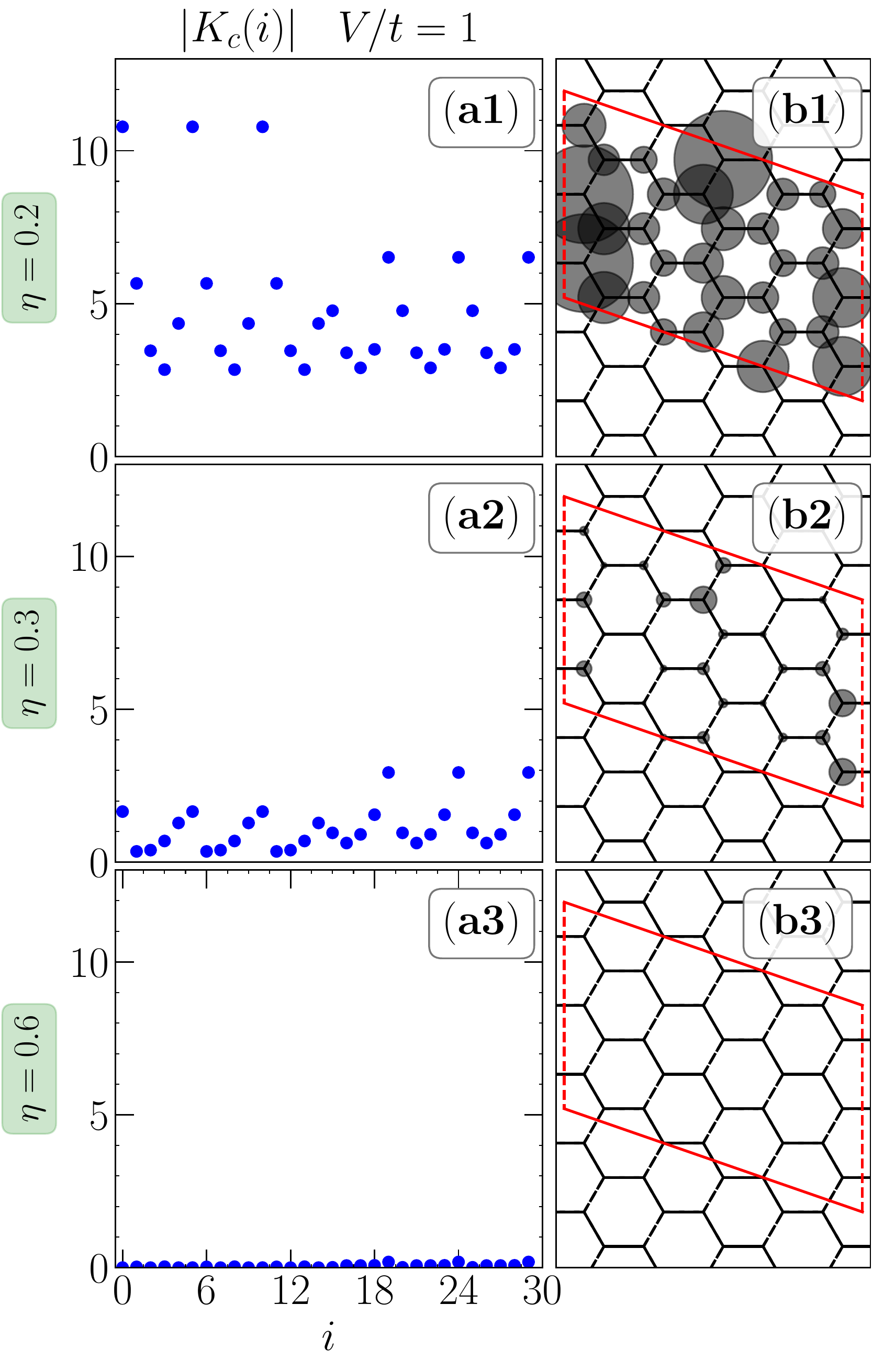}
\caption{Similar to Fig.~\ref{fig:Kc_eta}, but with fixed interactions $V/t =1$ and various dimerizations $\eta$, as marked.}
\label{fig:Kc_V}
\end{figure}

Two points are important to emphasize. While corner modes are essentially localized, these will always manifest a profile in real space, which poses challenges to its verification in small clusters (amenable to ED calculations). Second, a particularity that appears even in the $V=0$-limit of the dimerized Haldane-Hubbard model highlighted in Ref.~\cite{Wang2021} is that by studying semi-infinite ribbons, protected in-gap modes for $0<\eta<0.5$ are only observed in the case one employs an OBC cut across the $t_{1s}$ bonds (or strong bonds for $\eta < 1$). We follow a similar prescription here, defining mixed (or cylindrical) boundary conditions as schematically represented in Figs.~\ref{fig:Kc_eta} and \ref{fig:Kc_V} with the solid (PBC) and dashed (OBC) lines.

The transition from an interacting HOTI to the trivial Mott insulator under the scope of localized modes is shown in Fig.~\ref{fig:Kc_eta}, for a fixed $\eta=0.4$ and increasing interactions. It becomes apparent that the large compressibility at certain sites is quickly suppressed once charge-ordering, characteristic of the trivial Mott insulator, appears. Similarly, in Fig.~\ref{fig:Kc_V}, one can observe the interacting HOTI to CI transition with fixed interactions $V/t=1$ and increasing $\eta$.

\section{Summary and outlook}
Using exact calculations in small clusters, we show that a variant of the Haldane-Hubbard model displays a rich phase diagram, including unequivocal evidence of localized modes characteristic of high-order topology and the manifestation of a topological Mott insulator. The fundamental ingredient is the ${\cal C}_3$-symmetry breaking dimerization of the hoppings, which allows the topological and Mott transitions to be dissociated in this model. In doing so, the transition to a charge-ordered phase turns continuous (in opposition to first-order), whose universality class reflects the symmetry breaking of the CDW state.

\begin{figure*}[t!]
\centering
\includegraphics[width=2\columnwidth]{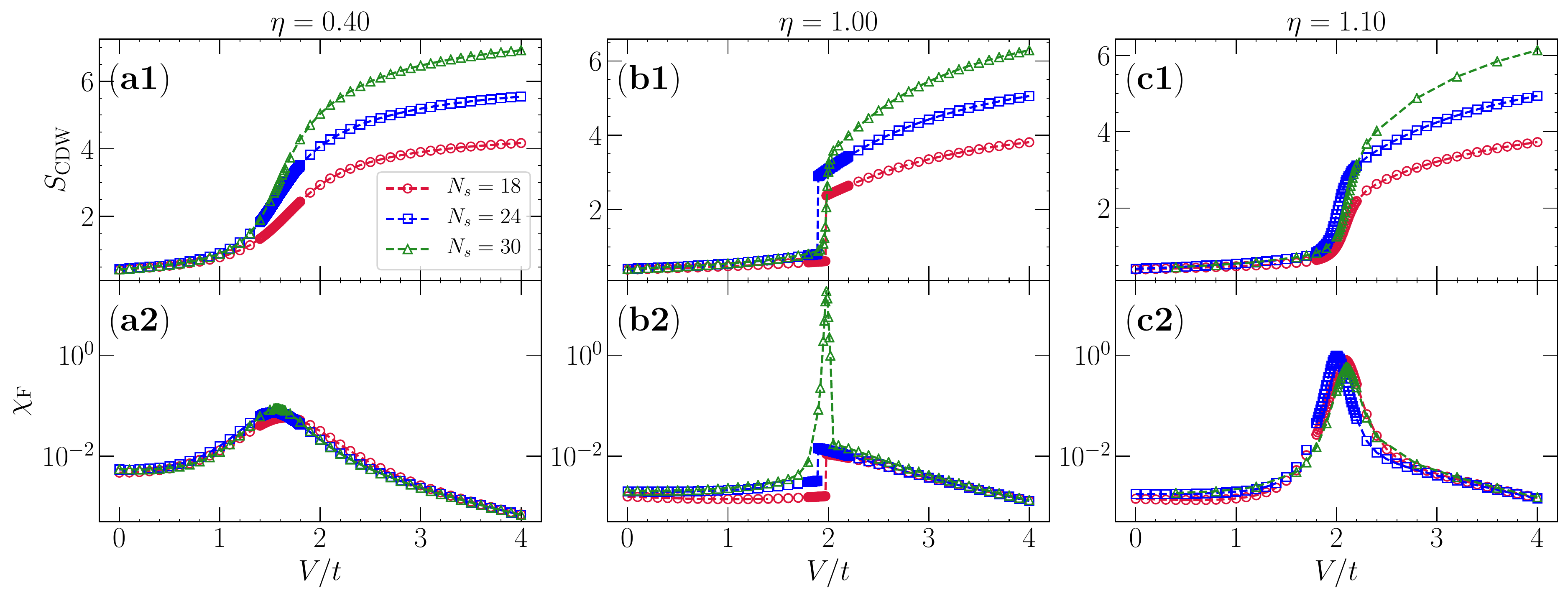}
\caption{(a1, b1, c1) CDW structure factor dependence with the interactions $V$ and different $\eta$'s [$\eta=0.4, 1$ and 1.1, respectively], contrasting the different system sizes we employ under PBCs. (a2, b2, c2) shows the same dependence but for fidelity susceptibility $\chi_F$. Discontinuities in both quantities are only seen in the homogeneous case $\eta=1$, signaling the Mott transition to be first-order.}
\label{fig:Scdw_N}
\end{figure*}

This modification in the hoppings in the Haldane model was originally introduced in Ref.~\cite{Wang2021} as a way to unveil localized corner modes characteristic of second-order topology and is a common way to construct high-order topological insulators~\cite{benalcazar2017quantized, schindler2018higher}. We show that the inclusion of interactions leads to a ground state adiabatically connected to it, generalizing the HOTI to the many-body realm. While exact calculations allow sufficient evidence for this characterization, the smallness of the clusters amenable to calculations and the fact the gapless charge excitations are not localized on a single site makes finite-size effects potentially relevant. Studying this model in a semi-infinite ribbon geometry, which, depending on how the OBCs are introduced, can harbor in-gap states in the non-interacting limit, is friendly to other techniques, including the infinite density matrix renormalization group~\cite{White92, Kjall13}. We envision the characterization of the HOTI phase, especially its boundaries to either the CI or the trivial Mott insulator, to be particularly sharp owing to the mitigated finite-size effects. We leave such an investigation to future studies.

\begin{acknowledgments}
We acknowledge support from the National Natural Science Foundation of China (NSFC) Grant No. NSAF-U1930402; H.-Q.~L. is supported by the NSFC Grant No.   ~12088101; R.M. acknowledges NSFC Grants No.~11974039 and No.~12050410263, No.~12111530010, and No.~12222401. Computations were performed on the Tianhe-2JK at the Beijing Computational Science Research Center.
\end{acknowledgments}

\appendix

\section{Finite-size effects in the Mott transition} \label{sec:FSE}
The main text shows the fidelity susceptibility and the charge-density-wave structure factor for cuts in the phase diagram and a single system size, $N_s =24$ [Fig.~\ref{fig:Fig2_24A_cuts}]. Figure~\ref{fig:Scdw_N} generalizes those results for the different system sizes studied. The general conclusions hold: discontinuities in $\chi_F$ and $S_{\rm CDW}$ are obtained in the $\eta =1$ (homogeneous) case, associated with the simultaneous Mott and topological transitions, whereas other values of $\eta$ show behavior typical of second-order phase transitions for the charge-ordered transition.

\bibliography{references}

\end{document}